\documentclass[sn-mathphys-num]{sn-jnl}

\usepackage{graphicx}%
\usepackage{multirow}%
\usepackage{amsmath,amssymb,amsfonts}%
\usepackage{amsthm}%
\usepackage{mathrsfs}%
\usepackage[title]{appendix}%
\usepackage{xcolor}%
\usepackage{textcomp}%
\usepackage{manyfoot}%
\usepackage{booktabs}%
\usepackage{algorithm}%
\usepackage{algorithmicx}%
\usepackage{algpseudocode}%
\usepackage{listings}%
\usepackage{orcidlink}%

\theoremstyle{thmstyleone}%
\theoremstyle{thmstyletwo}%

\theoremstyle{thmstylethree}%

\begin{document}

\title[Shear Viscosity of Collider-Produced QCD Matter I: AMY Formalism vs. A Modified Relaxation Time Approximation in 0-flavor SU(3) Theory]{Shear Viscosity of Collider-Produced QCD Matter I: AMY Formalism vs. A Modified Relaxation Time Approximation in 0-flavor SU(3) Theory}


\author[1]{\fnm{Noah M.} \sur{MacKay}\,\orcidlink{0000-0001-6625-2321}}\email{noah.mackay@uni-potsdam.de}

\affil[1]{\orgdiv{Institute of Physics and Astronomy}, \orgname{Universit\"at Potsdam}, \orgaddress{\street{Karl-Liebknecht-Stra\ss e 24/25}, \postcode{14476} \city{Potsdam}, \country{Germany}}}


\abstract{The AMY formalism is widely used to describe the transport coefficients of asymptotically hot and dense QCD matter, such as shear viscosity $\eta$. In literature prior to AMY, the viscosity of an asymptotically hot QCD plasma was expressed by a $q^2$ momentum transfer-weighted relaxation time approximation. Recent studies that compared numerical transport calculations and analytical expressions for $\eta$ demonstrated that asymptotically high temperatures and densities induce anisotropic scatterings, which are exhibited in the quark-gluon plasma produced by relativistic heavy ion collisions. In these studies, the QGP was treated as a Maxwell-Boltzmann-distributed gluon gas with added (anti-)quark degrees of freedom. One such method used in the comparison was the ``modified'' $q^2$ transport-weighted RTA. In this study, a comparison between the AMY formalism (both numerical calculations and next-leading-log expression) and the modified RTA expression for $\eta$ is made in 0-flavor SU(3) theory for collider-produced QGP. The comparison between numerical AMY calculations and the modified RTA method shows perfect agreement under the temperatures relevant for collider-produced QGP. Additionally, AMY is compared with the Chapman-Enskog method, which is well understood to better describe anisotropic collider-produced QGP.}




\maketitle

\section{Introduction}\label{sec1}

Relativistic heavy ion collisions produced at the Relativistic Heavy Ion Collider (RHIC) and at the Large Hadron Collider (LHC) create a form of high-temperature and high-density QCD matter, i.e., a quark-gluon plasma (QGP) containing (anti-)quark and gluon degrees of freedom \cite{J.Adams:STAR, PHENIX:2004vcz}. The partons are quasi-free in the plasma, where one can treat the QGP as an ideal thermal gas at temperatures much greater than the Hagedorn temperature: $T_H\simeq150$ MeV \cite{hage}.  It is suggested by past comparisons \cite{Romatschke:2007mq, Song:2008si, ALICE:2010suc} between experimental anisotropic flow measurements and hydrodynamic models (also kinetic transport models \cite{Lin:2001zk, Xu:2007jv, Xu:2008av, Ferini:2008he}) that the QGP behaves like a near-perfect fluid with a very small shear-viscosity to entropy-density ratio $\eta/s$ near the KSS limit of $1/(4\pi)$ \cite{Kovtun:2004de}. In this report, natural units are used, i.e., $\hbar=k_B=c=1$.

In numerical transport models, parton interactions are typically represented by their interaction cross section(s), including the magnitude and angular distribution, which then determine plasma properties such as the shear viscosity $\eta$. While hydrodynamic models treat the $\eta/s$ ratio (including its temperature dependence) as an input parameter, transport model calculations can extract calculations of QGP shear viscosity and $\eta/s$ after applying the relation between the parton cross section(s) and $\eta$ \cite{Xu:2007jv, Ferini:2008he, Xu:2007ns, Xu:2011fi}. The analytical analogue to numerical transport models is kinetic theory. For a two-body, i.e., $2\leftrightarrow 2$ collision, the collision kernel $\mathcal{C}[f]$ from the integro-differential Boltzmann equation, $p^\mu\partial_\mu f(\vec{p})=\mathcal{C}[f]$, depends on the the specific distribution function for each of the two bodies before the collision $f_{a,b}$ and after the collision $f_{c,d}$, where $f_i=f(\vec{p}_i)$; the factor $(1+\beta f_i)$ for each distribution function, where $\beta=1$ for Bose-Einstein (BE), $-1$ for Fermi-Dirac (FD), and $0$ for Maxwell-Boltzmann (MB); the relative motions between the two bodies, and the differential cross section of their shared interaction. Distinctive methods for calculating the collision kernel, either simply or explicitly, bring rise to various analytical expressions for the same plasma property, such as the shear viscosity.

For collider-produced QGP, it is repeatedly demonstrated that the Chapman-Enskog (CE) method \cite{ce, degroot} for expressing the shear viscosity (and $\eta/s$) agrees well with the numerical calculations under both isotropic and anisotropic scattering cases \cite{Plumari:2012ep, MacKay:2022uxo}. In both references, this method was commonly compared with a modified version of the relaxation time approximation (labeled as RTA$^*$), where the relaxation time $\tau$ utilizes the transport cross section $\sigma_{\mathrm{tr}}=\int d\sigma\,\sin^2\theta$ \cite{Molnar:2001ux} rather than the total cross section $\sigma=\int d\sigma$ per the traditional Anderson-Witting expression \cite{Anderson:1974nyl, Wiranata:2012br}. This is to account for the $q^2$ momentum transfer that the viscosity calculations also depend on. In Ref. \cite{MacKay:2022uxo}, in addition to RTA$^*$, the hydrodynamic Israel-Stewart \cite{Israel:1963, Stewart:1971, Huovinen:2008te} and Navier-Stokes \cite{degroot} methods were also compared. These methods, used for comparison with numerical calculations, inherently relied on massless MB statistics when solving the collision kernel, as well as a one-component simplification of the QGP as a gluon gas with added $N_f$-flavor quark degrees of freedom \cite{Plumari:2012ep, MacKay:2022uxo, Lin:2014tya, Zhang:1997ej, Wang:2021owa}. Therefore, the QGP was described by a differential cross section based on the perturbative QCD (pQCD) gluon-gluon amplitude (to be seen in Ref. \cite{Arnold:2003zc}, Appendix A).

As demonstrated in Ref. \cite{mackay1}, the temperature of the QGP influences the anisotropy of the scatterings. At higher temperatures, the shear viscosity is enhanced via the frequency of anisotropic scatterings. As a result, isotropic scatterings are more frequent at lower temperatures, albeit unphysically low for the QGP. For asymptotically high temperatures, the formalism framed by Arnold, Moore and Yaffe \cite{Arnold:2003zc, Arnold:2000dr} is widely used to express the shear viscosity and other transport coefficients of the QGP. What contributes to its reliable application is the formalism's ability to linearize the collision kernel for both $2\leftrightarrow2$ scatterings and $1\leftrightarrow2$ splitting/joining events, and the explicit useage of QFT and ``quantum'' statistics, i.e., BE and FD statistics for gluons and (anti-)quarks respectively with relevant degrees of freedom. These linearization calculations contribute to viscosity calculations under any number of quark flavors (up to 6, e.g., the up, down, strange, charm, top, and bottom flavors) and under any gauge symmetry group, such as the conventional SU(3) group for QCD. In this report, I will use the acronym ``AMY'' to describe the formalism from Arnold, Moore and Yaffe -- and the authors themselves -- provided in Refs. \cite{Arnold:2003zc, Arnold:2000dr}.

 However, in reports prior to AMY that analyzed the viscosity of an asymptotically hot plasma (i.e., a plasma with small $q^2$ momentum transfer) \cite{Thoma:1991em, Heiselberg:1994vy, Baym:1990uj, Baym:1991qf}, the relaxation time approximation was used. AMY acknowledges in the introduction section in Ref. \cite{Arnold:2003zc} that the so-called ``leading-log calculations'' are improvements to the previous RTA approach. However, there is a question whether AMY, despite its novelty and robustness, has an inherent RTA$^*$ behavior under a certain temperature region, especially in 0-flavor SU(3) theory (i.e., a gluon gas simplification of the QGP). This is the goal of this study, to compare the AMY formalism with RTA$^*$, using the exact pQCD gluon-gluon amplitude for both methods. Any discrepancy between quantum statistical calculations through AMY and MB statistical calculations through RTA$^*$ is negligible, as quantum statistics approximate to MB statistics under high temperatures and densities \cite{mackay1}. A secondary comparison can also be made, which compares the AMY and RTA$^*$ methods with the CE method, which is well understood to better describe anisotropic, i.e., high-temperature, collider-produced QGP. 

Asympotically high temperatures enable ultra-relativistic mechanics in the QGP, where the rest mass of partons is negelected due to $T\gg m$. In addition, the running gauge coupling $g$ is small due to asymptotic freedom \cite{Gross:1973id}, allowing the scattering amplitude $|\mathcal{M}|^2$ of parton interactions to be inserted into the differential cross section, where $d\sigma/d\Omega\propto|\mathcal{M}|^2$. With such differential cross sections, we can extract the shear viscosity and make comparisons.

\section{Methods}

\subsection{The Gluon-Gluon Interaction}

In previous studies \cite{Xu:2008av, Ferini:2008he, Plumari:2012ep, MacKay:2022uxo,  Molnar:2001ux, Zhang:1997ej, Wang:2021owa,  Lin:2004en}, numerical transport models, such as the ZPC parton cascade and a multi-phase transport (AMPT) model, utilized a cross section that is based on pQCD gluon-gluon scatterings. In this study, the pQCD gluon-gluon scattering amplitude \cite{Arnold:2003zc} is explicitly used, which is readily presented with the Debye screening mass $m_D$ in divergent channels as
\begin{equation} \label{pqcdgg}
|\mathcal{M}|^2=16C_A^2d_Ag^4\left(3-\frac{su}{(t-m_D^2)^2}-\frac{st}{(u-m_D^2)^2}-\frac{tu}{s^2} \right).
\end{equation}
In SU(3), $C_A=3$ and $d_A=8$ \cite{Arnold:2003zc}, which defines the SU(3) gluon-gluon matrix as follows (also to be seen in Ref. \cite{schwartz}):
\begin{equation} \label{pqcdgg}
|\mathcal{M}|^2=1152g^4\left(3-\frac{su}{(t-m_D^2)^2}-\frac{st}{(u-m_D^2)^2}-\frac{tu}{s^2} \right).
\end{equation}
In the above, $s$, $t$ and $u$ are the so-called Mandelstam variables from Feynman diagram calculations (${s}$ relates to the square of the center-of-momentum, or CM, energy; $t$ and $u$ both relate to the square of the momentum transfer). In some literature, the Mandelstam variables are given the hat notation $\hat{s},~\hat{t},~\hat{u}$; in this report I will not use the hat notation. As the external gluon lines are massless, $u=-t-s$; this defines Eq. (\ref{pqcdgg}) as purely dependent on the $s$- and $t$-Mandelstam variables.

In addition to $m_D$, $g$ is the QCD gauge coupling. As a running parameter, the QCD gauge coupling can be expressed as a function of $T$ \cite{Csernai:2006zz}:
\begin{equation}\label{rung}
\frac{1}{g^2(T)}=\frac{9}{8\pi^2}\ln\left(\frac{T}{\Lambda_T}\right)+\frac{4}{9\pi^2}\ln\left[2\ln\left(\frac{T}{\Lambda_T} \right) \right],
\end{equation}
where $\Lambda_T=30$ MeV, and one can define $g=\left[1/g^2(T) \right]^{-1/2}$. The Debye screening mass is defined to be proportional to the product $gT$ in $N_f$-flavor SU(3) theory \cite{Arnold:2003zc}:
\begin{equation}\label{debye}
m_D^2=\frac{1}{3}\left(3+\frac{1}{2}N_f \right)g^2T^2.
\end{equation}
For $N_f=0$, $m_D=gT$ straight-forwardly. As $g$ is a running parameter via Eq. (\ref{rung}), the Debye mass is also a running parameter.

Eq. (\ref{pqcdgg}) is readily summed over all spin-color states; however, it is not averaged by the incoming gluons' degrees of freedom. As gluons have 2 spin states and 8 gluon numbers, the averaging factor is $(2\times8)^2=256$. Thus, we have an averaged $gg\leftrightarrow gg$ amplitude, with $u=-t-s$:
\begin{equation} \label{glueamp}
\langle |\mathcal{M}|^2\rangle=\frac{9g^4}{2}\left(3+\frac{s(t+s)}{(t-m_D^2)^2}-\frac{st}{(t+s+m_D^2)^2}+\frac{t(t+s)}{s^2} \right).
\end{equation}

Provided in Ref. \cite{griffiths}, a mass-independent differential cross section $d\sigma/d\Omega$ in the CM frame is defined in terms of the spin-color averaged amplitude $\langle|\mathcal{M}|^2\rangle$. Based on the Mandelstam variables $s$ and $t$, the differential cross section is proportional to $1/s^2$ as it is differentiated by the momentum transfer variable $t$: 
\begin{equation}
\frac{d\sigma}{dt}=\frac{S}{16\pi s^2}\langle|\mathcal{M}|^2\rangle.
\end{equation}
Here,  $S$ is a statistical factor that corrects the over-counting of identical final-state particles. In the most general case for $n$ identical final-state particles of $m$ different species,
\begin{equation}
S=\prod_{i=1}^m\frac{1}{n_i!}.
\end{equation}
In the case of no identical particles in the final state, $S=1$. In our case for $gg\leftrightarrow gg$ ($n=2$ and $m=1$), $S=1/2$. Therefore, 
\begin{equation} \label{realdsdt}
\begin{split}
\frac{d\sigma}{dt}&=\frac{1}{32\pi s^2}\langle|\mathcal{M}|^2\rangle\\
&=\frac{9g^4}{64\pi s^2}\left(3+\frac{s(t+s)}{(t-m_D^2)^2}-\frac{st}{(t+s+m_D^2)^2}+\frac{t(t+s)}{s^2} \right).
\end{split}
\end{equation}

\subsection{Shear Viscosity} \label{shear}

Allowing a MB-distributed system with no chemical potential to be slightly perturbed out of equilibrium, the distribution function in the Boltzmann equation $f(\vec{p})$ can be expressed as an expansion containing the local equilibrium distribution and a small deviation: 
\begin{equation} \label{lindist}
f(\vec{p})=f_0(\vec{p})[1+\phi].
\end{equation}
Here, $f_0(\vec{p})=\exp(-U_\nu p^\nu/T)$ is the local equilibrium distribution with $U_\nu p^\nu=p^0\equiv\sqrt{|\vec{p}|^{2}+m^2}$ under the metric $g_{\mu\nu}=\mathrm{diag}(+1,-1,-1,-1)$ ($U^\mu$ is the 4-flow velocity whereby $U_\mu U^\mu=1$, and $p^\mu$ is the 4-momentum of a particle). Throughout the derivations, ultra-relativistic (i.e., massless) particles, such as the QGP partons, will be considered. In addition, 
\begin{equation}
\phi=\varepsilon_0+\varepsilon_\mu p^\mu+\varepsilon_{\mu\nu}p^\mu p^\nu+\dots
\end{equation}
is a polynomial deviation function that satisfies $|\phi|\ll1$, where the choice of the highest order in the polynomial (i.e., the highest moment) fine-tunes the evaluation of the collision kernel $\mathcal{C}[f]$. Through this linearization procedure in the distribution function and in the deviation polynomial, the kernel is linearized as an operator acting on $\phi$ and using the local equilibrium distribution:
\begin{equation} \label{linboltz}
p^\mu\partial_\mu f_0(\vec{p})=-f_0(\vec{p})\mathcal{L}[\phi].
\end{equation}

This is where solving the Boltzmann equation can go on different paths, based on the preferred moment chosen to linearize the collision kernel. This is how we have the various approaches -- namely the relaxation time approximation and the Israel-Stewart, Navier-Stokes and Chapman-Enskog methods -- that each assigns a specific Ansatz for defining $\phi$ and obtaining $\mathcal{L}[\phi]$. However, all methods have a common step in solving the left-hand side of Eq. (\ref{linboltz}). Using $f_0(\vec{p})=\exp(-U_\nu p^\nu/T)$, the left-hand side of Eq. (\ref{linboltz}) comes to be
\begin{equation}
p^\mu\partial_\mu f_0(\vec{p})=-f_0(\vec{p})\frac{p^\mu p^\nu}{T}\left(\partial_\mu U_\nu-\frac{1}{T}U_\nu\partial_\mu T \right),
\end{equation}
where the term $T^{-1}\partial_\mu T$ is defined by the conservation laws $\partial_\nu T^{\mu\nu}=0$ and $\partial_\mu S^\mu=0$ for the energy-momentum and entropy tensors, respectively. For the case of ultra-relativistic partons with a square of the sound velocity of $v_s^2\equiv dP/d\epsilon=1/3$ (c.f. Eqs. (42) and (43) in Ref. \cite{Chakraborty:2010fr}), one can obtain the following for the left-hand side of Eq. (\ref{linboltz}): 
 \begin{equation}
 \begin{split}
&p^\mu\partial_\mu f_0(\vec{p})=-f_0(\vec{p})\frac{p^\mu p^\nu}{2T}\left[\nabla_{\langle\mu}U_{\nu\rangle}+\frac{2}{3}g_{\mu\nu}\partial_\rho U^\rho \right],\\
&\mathrm{where}~~~\nabla_{\langle\mu}U_{\nu\rangle}=\left(D_\mu U_\nu+D_\nu U_\mu+\frac{2}{3}\Delta_{\mu\nu}\partial_\rho U^\rho \right),
\end{split}
\end{equation}
$D_\mu=\partial_\mu-U_\mu U^\nu\partial_\nu$, and $\Delta_{\mu\nu}=U_\mu U_\nu-g_{\mu\nu}$.

Shear viscosity $\eta$, in part with the bulk viscosity $\zeta$, comes from the dissipative part of the energy-momentum tensor $\pi_{\mu\nu}$, which is generally defined as a linear combination of a shearing and bulking factor \cite{Chakraborty:2010fr}:  
\begin{equation}
\pi_{\mu\nu}=\eta\nabla_{\langle\mu}U_{\nu\rangle}-\zeta\Delta_{\mu\nu}\partial_\rho U^\rho.
\end{equation}
The left-hand side of Eq. (\ref{linboltz}) can therefore be expressed in terms of the dissipative tensor and the viscosities. However, setting the condition that $\partial_\rho U^\rho=0$ in the local rest frame, only the shearing contribution takes precedence:
\begin{equation}
\frac{-p^\mu p^\nu}{2T}\frac{\pi_{\mu\nu}}{\eta}f_0(p)=-f_0(p)\mathcal{L}[\phi].
\end{equation}
From this revision of the Boltzmann equation, one can define the shear viscosity $\eta$ as inversely proportional to the linearized collision kernel, i.e., $\eta\propto\mathcal{L}[\phi]^{-1}$. Furthermore, as shear viscosity describes the thermodynamic flow of the off-diagonal shear stress tensor $\pi_{ij}$, the spacetime indices $\mu\nu$ only need to be the orthogonal 3-space indices $ij$:
\begin{equation} \label{viscdef}
\eta=\frac{1}{2T}\frac{\pi_{ij}p^i p^j}{\mathcal{L}[\phi]}.
\end{equation}

Now we define the shear stress tensor. The energy-momentum tensor in the local rest frame is ideally defined as $T_{\mu\nu}=(\epsilon+P)U_\mu U_\nu-Pg_{\mu\nu}$. As the energy density $\epsilon$ and ultra-relativistic pressure $P=\epsilon/3$ are defined as momentum-space integrals in kinetic theory, the tensor itself can be written as a momentum-space integral, under any distribution with quantum degrees of freedom $d_j$:
\begin{equation}
T^{\mu\nu}=d_j\int_{-\infty}^\infty \frac{d^3\vec{p}}{(2\pi)^3}\frac{p^\mu p^\nu}{p^0} f(\vec{p}).
\end{equation}
For a slightly perturbed system, we use the expansion of the distribution function provided as Eq. (\ref{lindist}) in the integral definition of $T^{\mu\nu}$. This is to define the tensor at local equilibrium and its small deviation, i.e., $T^{\mu\nu}= T^{\mu\nu}_{(0)}+\pi^{\mu\nu}$: 
\begin{equation}
T^{\mu\nu}=d_j\int_{-\infty}^\infty \frac{d^3\vec{p}}{(2\pi)^3}\frac{p^\mu p^\nu}{p^0} f_0(\vec{p})+d_j\int_{-\infty}^\infty \frac{d^3\vec{p}}{(2\pi)^3} \frac{p^\mu p^\nu}{p^0} f_0(\vec{p})|\phi|.
\end{equation}
Here, $\pi^{\mu\nu}$ is the dissipative part of the energy-momentum tensor, which is readily defined as an integral:
\begin{equation}
\begin{split}
\pi^{\mu\nu}=d_j\int_{-\infty}^\infty \frac{d^3\vec{p}}{(2\pi)^3} \frac{p^\mu p^\nu}{p^0} f_0(\vec{p})|\phi|.
\end{split}
\end{equation}
This integral expression is equivalent to Eq. (36) in Ref. \cite{Chakraborty:2010fr}, however the integral is summed over all relevant particle types in the gas. For a one-component mixture, such as the gluon gas simplification for the QGP, the summation is dropped, as readily defined.

 For Eq. (\ref{viscdef}), the spacetime indices of $\pi^{\mu\nu}$ only need to be the off-diagonal 3-space indicies $ij$. This defines, therefore, an integral definition for the shear viscosity using $|\phi|=2/15$:
\begin{equation} \label{visintmass}
\eta=\frac{d_j}{15T}\int_{-\infty}^\infty \frac{d^3\vec{p}}{(2\pi)^3}\frac{|\vec{p}|^4}{p^0}\frac{f_0(\vec{p})}{\mathcal{L}[\phi]},
\end{equation}
which is analogous to Eq. (39) in Ref. \cite{Chakraborty:2010fr}. Specifically for massless partons, $p^0=|\vec{p}|$ would be implied.

\subsubsection{The Relaxation Time Approximation}

RTA was first used by Bhatnagar, Gross and Krook \cite{Bhatnagar:1954zz}, which is designed to solve the linearized collision kernel by assigning a characteristic frequency at which collisions in local equilibrium occur. Its inverse, therefore, is characterized as the time between consecutive collisions in local equilibrium. In terms of the linearized kernel, an Ansatz can be made where the first moment, i.e., the first-order linear term of the $\phi$ polynomial is the particle energy in local equilibrium, $\varepsilon_\mu p^\mu=U_\mu p^\mu\equiv p^0$, and the characteristic frequency (defined as an integral via the linearized kernel) is factored in:
\begin{equation}
\mathcal{L}[\phi]=\frac{p^0}{\tau}.
\end{equation}
Here, $\tau=\omega^{-1}$ is the relaxation time, which is for now left undefined. Specifically  with $p^0=|\vec{p}|$, Eq. (\ref{visintmass}) is defined in the massless RTA case as
\begin{equation}\label{rta1}
\eta=d_j\frac{\tau}{15T}\int_{-\infty}^\infty \frac{d^3\vec{p}}{(2\pi)^3}|\vec{p}|^2f_0(\vec{p}).
\end{equation}

Eq. (\ref{rta1}) can be reduced even further for a simpler definition. Directly from statistical mechanics, the number density and the average of a given quantity $a$ under any distribution (with quantum degeneracy and in local equilibrium) are defined as follows:
\begin{equation}
\begin{split}
&n=d_j\int_{-\infty}^\infty \frac{d^3\vec{p}}{(2\pi)^3}f_0(\vec{p}),~~~\langle a\rangle=\frac{1}{Z_0}\int_{-\infty}^\infty d^3\vec{p}\, a(\vec{p})f_0(\vec{p}),\\
&~~~~~~~~~~~~~~~~Z_0\equiv\int_{-\infty}^\infty d^3\vec{p}\,f_0(\vec{p})=\frac{(2\pi)^3}{d_j}n.
\end{split}
\end{equation}
From these definitions, the massless RTA shear viscosity can be straight-forwardly redefined as
\begin{equation}\label{rta2}
\eta=\frac{\langle|\vec{p}|^2\rangle}{15T}n\tau.
\end{equation}
Under massless MB statistics, $\langle|\vec{p}|^2\rangle=12T^2$. Also under massless MB statistics, the energy density is defined as $\epsilon=3nT$, which is introduced through $\langle|\vec{p}|^2\rangle$:
\begin{equation}\label{rta3}
\eta=\frac{4}{15}\epsilon\tau.
\end{equation}

Eq. (\ref{rta3}) was previously obtained in the numerical Green-Kubo context in Ref. \cite{Wang:2021owa}, whereby the relaxation time $\tau$ was numerically calculated under a ZPC parton cascade model. In this provided context, $\tau$ is an integral expression; traditionally, the Anderson-Witting relaxation time \cite{Anderson:1974nyl, Wiranata:2012br} is used, which expresses $\tau$ in terms of the number density, an energy-independent total cross section, and the relative velocity between two particles in a two-distribution, i.e., thermal average:
\begin{equation}
\tau=\frac{1}{n\sigma\langle v_{\mathrm{rel}}\rangle}.
\end{equation} 
For massless particles, $\langle v_{\mathrm{rel}}\rangle=1$.

\subsubsection{Modified RTA (RTA$^*$)}

On general physical grounds, the viscosity also depends on the $q^2$ momentum transfer that the collisions produce on average \cite{Plumari:2012ep, MacKay:2022uxo}. This momentum transfer would therefore infleunce the effective measurement on the scattering cross section. In literature sometimes, the transport cross section $\sigma_{\mathrm{tr}}$ is considered in lieu of the total cross section in the Anderson-Witting formula, where \cite{Molnar:2001ux}
\begin{equation}
\sigma_{\mathrm{tr}}\equiv\int d\sigma\,\sin^2\theta=\int_{-s}^0 \frac{d\sigma}{dt}\left(-\frac{4t^2}{s^2}-\frac{4t}{s} \right)dt.
\end{equation}
The latter definition is in terms of the $s$- and $t$-Mandelstam variables, and $d\sigma/dt$ is the momentum-transfer differential cross section, such as Eq. (\ref{realdsdt}). 

To account for all values of $q^2$ momentum transfer in the transport cross section, it is placed in the thermal average to be evaluated with the relative velocity. In Ref. \cite{Plumari:2012ep}, the Anderson-Witting relaxation time was replaced by this new relaxation time: 
\begin{equation}
\tau^{*}=\frac{1}{n\langle \sigma_{\mathrm{tr}}v_{\mathrm{rel}}\rangle},
\end{equation}
where, for the case of massless partons \cite{Koch:1986ud},
\begin{equation} \label{vrelint}
\langle \sigma_{\mathrm{tr}} v_{\mathrm{rel}}\rangle= \frac{1}{16}\int_0^\infty\sigma_{\mathrm{tr}}\left[\tilde{u}^4K_1(\tilde{u}) \right]d\tilde{u},
\end{equation}
$K_n$ is the modified Bessel function of the second kind, and $\tilde{u}=\sqrt{{s}}/T$ is the integration variable. This leads to the definition of shear viscosity under a modified version of RTA, or RTA$^*$ \cite{Plumari:2012ep, MacKay:2022uxo}:
\begin{equation} \label{modrta}
\eta^{\mathrm{RTA^*}}=\frac{4}{5}\frac{T}{\langle \sigma_{\mathrm{tr}} v_{\mathrm{rel}}\rangle}.
\end{equation} 

\subsubsection{$\eta$ in pQCD} \label{amydisc}

The AMY formalism is widely used to express the shear viscosity of a QGP in pQCD \cite{Arnold:2003zc}. In this regime, the plasma temperature is asymptotically high and QFT is applied to the Boltzmann equation. In reports prior to AMY that analyzed the viscosity of an asymptotically hot plasma \cite{Arnold:2000dr, Thoma:1991em, Heiselberg:1994vy, Baym:1990uj, Baym:1991qf}, RTA with $q^2$ momentum transfer-weighting was used. Specifically in Refs. \cite{Thoma:1991em, Heiselberg:1994vy, Baym:1990uj, Baym:1991qf}, the obtained relaxation rate of a weak interacting QCD plasma, $\tau^{-1}$, is proportional to the inverse log of the QCD structure constant $\alpha_s$: 
\begin{equation} \label{pqcdtau}
\frac{1}{\tau}\propto\alpha_s^2\ln\left(1/\alpha_s\right)T+\mathcal{O}(\alpha_s^2)\propto g^4T\ln\left(1/g\right),
\end{equation}
where $g=\sqrt{4\pi\alpha_s}$. If one applies Eq. (\ref{pqcdtau}) in place of $1/\tau$ in Eq. (\ref{rta3}) with $\epsilon\propto T^4$, the obtained expression for viscosity is the following: 
\begin{equation}
\eta=\kappa\frac{T^3}{g^4\ln(1/g)}.
\end{equation}
 Here, $\kappa$ is a proportionality constant. This expression for shear viscosity is provided as Eq. (1.14) in Ref. \cite{Arnold:2000dr}. Table 2 in Ref. \cite{Arnold:2000dr} offers a list of specific values for $\kappa$ depending on number of quark flavors in SU(3) theory.
 
Should one follow this formula directly as provided with Eq. (\ref{rung}) as $g$, a negative curve of $\eta/\kappa$ is produced for all temperatures. This led AMY in Ref. \cite{Arnold:2003zc} to use expansion of inverse logs to provide a correction. Firstly, AMY considers a collision kernel $\mathcal{C}[f]$ that includes both $2\leftrightarrow2$ and $1\leftrightarrow2$ interactions. The $2\leftrightarrow2$ kernel is generally defined for the collision $a+b\rightarrow c+d$, whether elastic or inelastic, and under quantum statistics:
\begin{equation} \label{amykernel}
\begin{split}
\mathcal{C}[f]=\frac{1}{4|\vec{p}_a|\nu_a}\sum_{bcd}\int& |\mathcal{M}|^2(2\pi)^4\frac{\delta^4(p_a+p_b-p_c-p_d)}{2|\vec{p}_b|2|\vec{p}_c|2|\vec{p}_d|}\frac{d^3\vec{p}_b}{(2\pi)^3}\frac{d^3\vec{p}_c}{(2\pi)^3}\frac{d^3\vec{p}_d}{(2\pi)^3}\\
&\times\Big( f_af_b[1\pm f_c][1\pm f_d]-f_cf_d[1\pm f_a][1\pm f_b]\Big).
\end{split}
\end{equation}
Here, $f_i=f(\vec{p}_i)$ for the particle types $a,~b,~c$ and $d$; the $a+b\rightarrow c+d$ interaction is depicted in the summed (but not averaged) scattering amplitude $|\mathcal{M}|^2$. The kernel itself is summed only over the types $b,~c,~d$. In addition, $\nu_a$ is the spin-color states of particle type $a$, i.e., 6 for (anti-)quarks and 16 for gluons. In 0-flavor SU(3) theory, i.e., for the pure gluon gas, all particles are of the same species and the summation is dropped. In addition, BE statistics are utilized, where only the Bose enhancement $[1+f_i]$ is applied in $\mathcal{C}[f]$; and $|\mathcal{M}|^2$ is the unaveraged gluon-gluon amplitude (Eq. [\ref{pqcdgg}]).

AMY defines the next-leading-log viscosity as follows (c.f. Eq. (2.25c) in Ref. \cite{Arnold:2003zc}):
\begin{equation} \label{appdxnll}
\eta^{\mathrm{NLL}}=\frac{2}{15}Q_{\mathrm{max}},
\end{equation}
where $Q_{\mathrm{max}}$ maximizes the expectation of the inverse linearized collision kernel, provided as a sum in Eq. (4.11) in Ref. \cite{Arnold:2003zc}. For the first order ($n=1$) term, one has $Q_{\mathrm{max}}\propto \mathcal{A}^{-1}/(\ln(\mu_*/m_D))$, where 
\begin{equation}
\mathcal{A}=\lim_{w\rightarrow0}\left\{-w\frac{\partial}{\partial w}\left[\frac{\mathcal{C}(w)}{g^4T}\right]\right\}
\end{equation}
with $w=m_D/T$.

The factor $\mathcal{A}$ has the implied minimizing condition of $m_D/T\rightarrow0$ on Eq. (\ref{amykernel}). This was done via a variational method of basis functions acting on the matrix. After minimization, the kernel is then subjugated in a log expansion; as the viscosity is inversely proportional to the linearized kernel, the log expansion acting on the linearized kernel acts as an inverse log expansion on the viscosity. This defines a geometric sum for the viscosity; only the first term is considered for the ``next-leading-log'' expression of the viscosity:
\begin{equation} \label{nll}
\eta^{\mathrm{NLL}}=\frac{T^3}{g^4}\left[\frac{\eta_1}{\ln(\mu_*/m_D)} \right].
\end{equation}
Here, $\mu_*/T$ is an arbitrary scale (so that $\mu_*/m_D\propto1/g$ rather than $=1/g$), whose value is specific for number of quark flavors and gauge symmetry (see Table 1 in Ref. \cite{Arnold:2003zc}). Obtaining $\eta_1$ depends on the calculation of the linearized collision kernel $\mathcal{C}(m_D/T)$, which in turn also depends on the number of quark flavors and the gauge symmetry implied in $|\mathcal{M}|^2$. For 0-flavor SU(3), AMY calculates $\eta_1=27.126$ with the scaling $\mu_*/T=2.765$ (also provided in Ref. \cite{Ghiglieri:2018dib}).

\begin{figure}[h!]
\centering
\includegraphics[width=125mm]{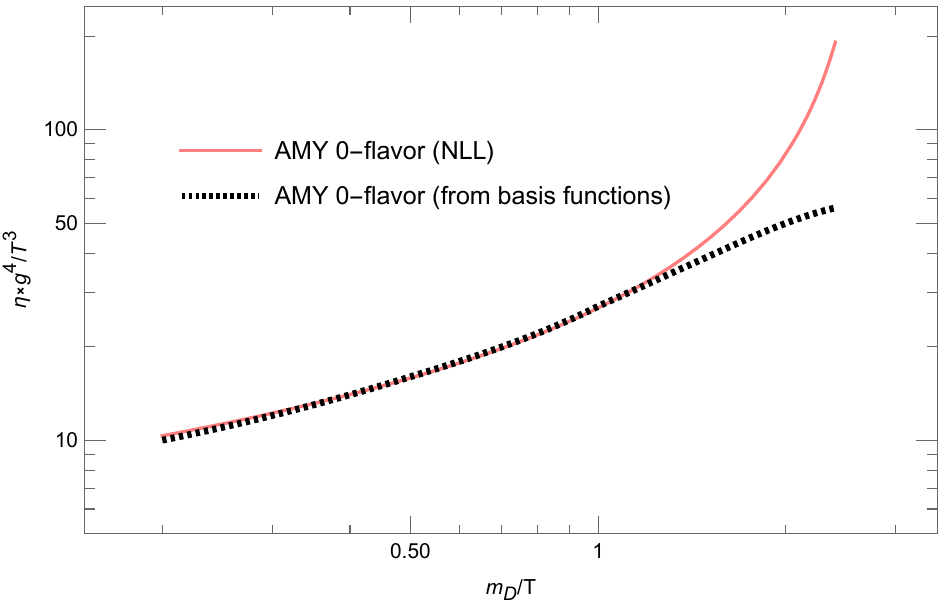}
\caption{\label{fig:log_basis} 0-flavor SU(3) AMY curves for $\eta\times g^4/T^3$ vs. $m_D/T$; the NLL log expression is drawn as the red solid curve, and the basis function calculations are interpolated as the black dashed curve.}
\end{figure}

Note that Eq. (\ref{nll}) is a ``high-temperature,'' i.e., $m_D/T\rightarrow0$ approximation to the exact numerical calculations from the basis functions. As shown in Figure \ref{fig:log_basis} for 0-flavor SU(3) theory, the mutual agreement between Eq. (\ref{nll}) and the exact calculations is present under $m_D/T\leq1$, as per design. However, under arguments of $m_D/T>1$, the inverse-log curve begins to diverge away from the basis function curve. The 0-flavor SU(3) basis function curve is readily offered in Figure 2 in Ref. \cite{Arnold:2003zc}. Alternatively, one can avoid the basis function calculations by deducing a sufficient amount of data points along the 0-flavor curve, and from the dataset interpolate a function to smoothen the list-line plot into a flowing curve. In Figure \ref{fig:log_basis}, the interpolation alternative was used.
 
Divergence in Eq. (\ref{nll}) occurs as $\mu_*/m_D\rightarrow1$, i.e., as $m_D/T\rightarrow\mu_*/T$. For 0-flavor SU(3), $m_D/T\equiv g$, meaning divergence occurs as $g\rightarrow2.765$. This corresponds to a threshold temperature of $T=74.5$ MeV via Eq. (\ref{rung}), which is much below the Hagedorn temperature. For collider-produced QGP, where the temperature window is on average between $T=550$ MeV and $T=150$ MeV \cite{MacKay:2022uxo, Lin:2014tya}, this relates to a relevant x-axis region of $m_D/T\in[1.56, 2.06]$. In Figure \ref{fig:log_basis}, it is the region along the x-axis where the NLL formula begins to diverge from the basis function curve.

Therefore, for a proper comparison between the AMY and RTA$^*$ viscosities for a collider-produced QGP, the numerical calculations via the basis functions will be compared with RTA$^*$. Eq. (\ref{nll}) will also be provided for comparison in the x-region of $m_D/T\leq1$. 

\section{$\eta\times g^4/T^3$ vs. $m_D/T$: AMY vs. RTA$^*$} \label{comp}

Using Eq. (\ref{realdsdt}), the transport cross section for the pQCD gluon-gluon interaction is defined as follows, given the variable combinations $w=m_D/T$ and $\tilde{u}=\sqrt{s}/T$:
\begin{equation}
\begin{split}
\sigma_{\mathrm{tr}}&=\frac{9g^4}{8\pi T^2\tilde{u}^2}\left[-3\left(\frac{w^2}{\tilde{u}^2}+\frac{34}{45} \right)+\left(\frac{w^2}{\tilde{u}^2}+1\right)\left(\frac{3w^2}{\tilde{u}^2}+1\right)\ln\left(1+\frac{\tilde{u}^2}{w^2}\right) \right]\\
&\equiv\frac{9g^4}{8\pi T^2\tilde{u}^2}H\left(\frac{w^2}{\tilde{u}^2} \right).
\end{split}
\end{equation}
The last line is to redefine the gluon-gluon transport cross section as a short-hand function of the x-axis variable $w$ from Figure \ref{fig:log_basis} and the integration variable $\tilde{u}$ from Eq. (\ref{vrelint}). Furthermore, one can reformulate Eq. (\ref{modrta}) to make it comparable to Eq. (\ref{nll}):
\begin{equation}
\begin{split}
\eta^{\mathrm{RTA^*}}=\frac{T^3}{g^4}\left[\frac{35.744}{\int_0^\infty H\left({w^2}/{\tilde{u}^2} \right)\cdot[\tilde{u}^2K_1(\tilde{u})]d\tilde{u} }\right].
\end{split}
\end{equation}

\begin{figure}[h!]
\centering
\includegraphics[width=125mm]{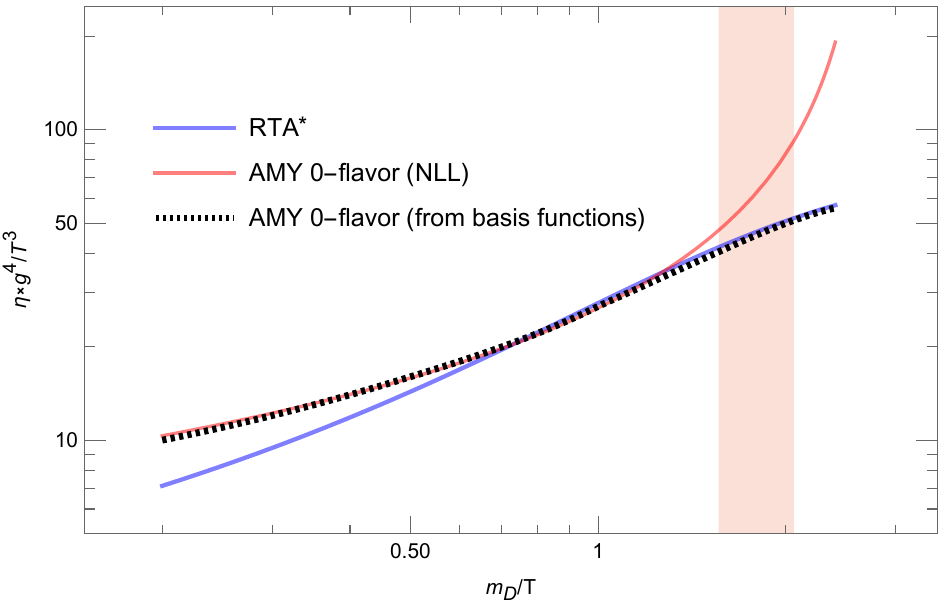}
\caption{\label{fig:rta_amy} $\eta\times g^4/T^3$ vs. $m_D/T$ under RTA$^*$ (blue solid) and 0-flavor SU(3) AMY formalism (red solid via NLL, black dashed from the basis functions). The pink region along the x-axis highlights the range of values of running coupling $g$ relevant to a collider-produced QGP.}
\end{figure}

Figure \ref{fig:rta_amy} depicts the log-log plot of $\eta\times g^4/T^3$ vs. $m_D/T$ under RTA$^*$ and 0-flavor SU(3) AMY formalism (both inverse-log expression and basis function calculations). As seen in the figure, small arguments of $m_D/T$ show that RTA$^*$ severely underestimates the highly anisotropic QGP viscosity. The same conclusion was made in Refs. \cite{Plumari:2012ep, MacKay:2022uxo}, which instead used a ``large $s$'' gluon-gluon amplitude based on Eq. (\ref{pqcdgg}). However, for $m_D/T>1$, the AMY basis function calculations for 0-flavor SU(3) agree well with the RTA$^*$ curve, especially within the relevant region of $m_D/T\equiv g$ that corresponds to gauge coupling values for a collider-produced QGP.

\section{Discussion} \label{disc}

To reiterate, the question proposed in Section \ref{sec1} was whether the 0-flavor SU(3) AMY formalism has an inherent $q^2$ momentum transfer-weighted RTA behavior under a certain temperature region. After the discussions and derivations provided in Section \ref{shear} -- for the shear viscosity under RTA$^*$ via Eq. (\ref{modrta}) and under AMY's inverse-log expression via Eq. (\ref{nll}) --, and after a brief discussion on AMY's exact calculations of $\eta$ using basis functions and how it compares with Eq. (\ref{nll}), Figure \ref{fig:rta_amy} displays the comparison between the AMY formalism and the RTA$^*$ expression.

The results of the comparison show that the AMY calculations from the basis functions mutually agree with RTA$^*$ for large $m_D/T$, while for small $m_D/T$ the numerical results agree well with the inverse-log expression. More importantly, RTA$^*$ and AMY's numerical results mutually agree in the relevant range of $m_D/T$ that is associated with gauge coupling values for a collider-produced QGP. This is only under the pretense that the QGP is simplified as a gluon gas, where $m_D/T=g$. 

\subsection{AMY and RTA$^*$ vs. CE}

\begin{figure}[h!]
\centering
\includegraphics[width=125mm]{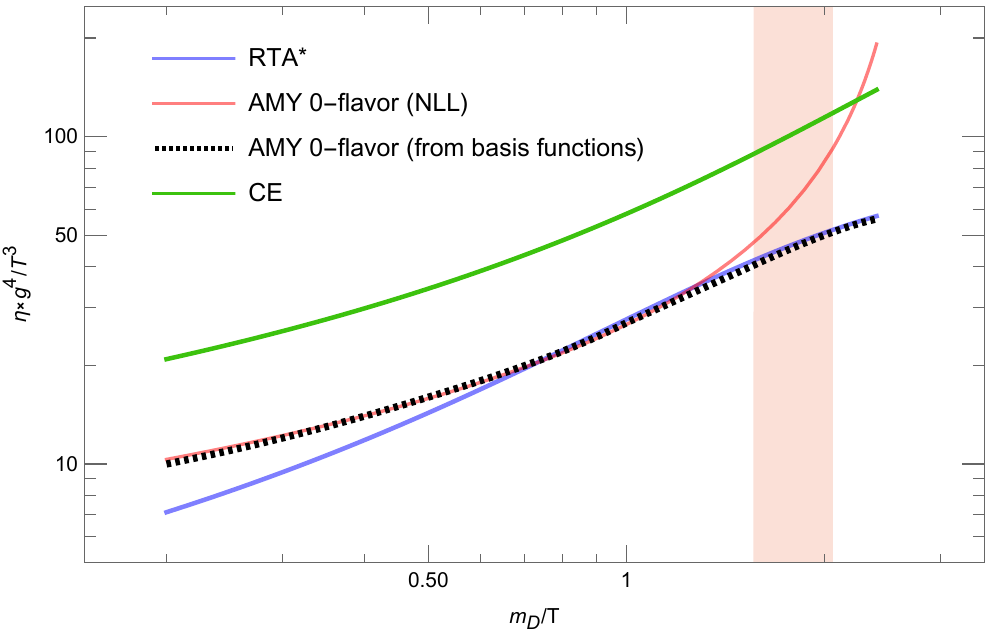}
\caption{\label{fig:rta_amy_ce} $\eta\times g^4/T^3$ vs. $m_D/T$ under RTA$^*$ (blue solid), AMY 0-flavor formalism (black dashed from basis functions, red solid via NLL) and CE (green solid). Like in Figure \ref{fig:rta_amy}, the pink region highlights the range of values of running coupling $g$ relevant to a collider-produced QGP.}
\end{figure}

Using a differential cross section based on Eq. (\ref{realdsdt}) for transport models, the Chapman-Enskog method is well understood to agree well with the numerical Green-Kubo calculations under both isotropic and anisotropic scatterings \cite{Plumari:2012ep, MacKay:2022uxo}. It was shown in both references that the CE method demonstrates an anisotropic enhancement of the shear viscosity, which correlates to high temperatures \cite{mackay1}. With this in consideration, a secondary comparison can be made between the AMY formalism (both numerical and analytical) and the RTA$^*$ and CE methods, which is shown in Figure \ref{fig:rta_amy_ce}. 

With the inclusion of the CE curve for $\eta\times g^4/T^3$ in the comparison, one can see that the CE curve is larger than the other comparable methods over all $m_D/T$, except for the inverse-log expression as it diverges towards $m_D/T\rightarrow2.765$. One can even deduce that the CE curve is greater than the numerical AMY calculations roughly by a factor of 2 over all $m_D/T$. As anisotropy is understood to enhance the shear viscosity, one can claim that the CE method is the preferrable method for an asymptotically hot QGP, at least in 0-flavor SU(3) theory.

\subsection{$N_f$-flavor AMY and $N$-Component RTA$^*$} \label{ncomp}

The reliability of the AMY formulism is contributed to the treatment of $N_f$-flavor quarks in the linearized collision kernel, i.e., the explicit utilization of FD statistics, Pauli blocking factors, and fermion kinematics via $|\mathcal{M}|^2$ in Eq. (\ref{amykernel}). In SU(3), the calculations of $\eta_1$ and the scale $\mu_*/T$ for $N_f>0$ are provided in Table \ref{tab:i}. 

\begin{table}[h!]
\centering
\begin{tabular}{lll}
\hline
$N_f>0$	 & $\eta_1$		& $\mu_*/T$   \\
\hline 
2		 & 86.470		& 2.954	\\
3		 & 106.66		& 2.957	\\
4		 & 122.96		& 2.954	\\
5		 & 136.38		& 2.947	\\
6		 & 147.63		& 2.940	\\
\hline
\end{tabular}
\caption{\label{tab:i} AMY calculations of $\eta_1$ and $\mu_*/T$ in $N_f$-flavor SU(3) theory. Flavor numbers provided range from 2 to 6.}
\end{table}

\begin{figure}[h!]
\centering
\includegraphics[width=125mm]{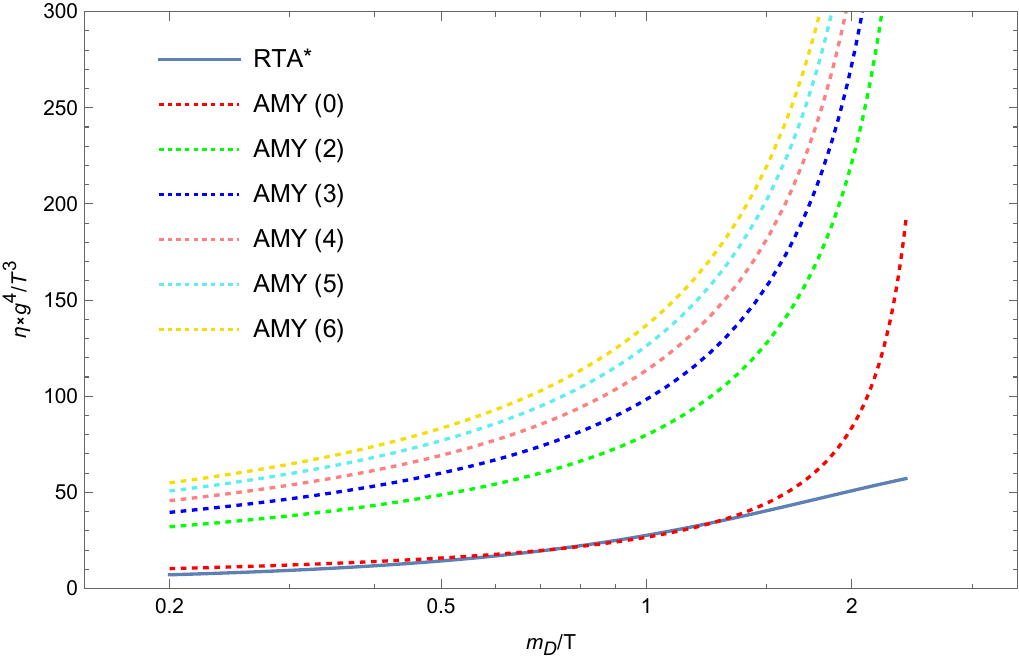}
\caption{\label{fig:rta_amy_large} RTA$^*$ versus $N_f$-flavor SU(3) AMY for $\eta\times g^4/T^3$ vs. $m_D/T$.}
\end{figure}

While Figure 2 of Ref. \cite{Arnold:2003zc} maps the $\eta\times g^4/T^3$ vs. $m_D/T$ for flavor numbers $N_f=0,~2$--$6$ via the basis function calculations, Figure \ref{fig:rta_amy_large} maps the quantity via the inverse-log expression (Eq. [\ref{nll}]) with the parameters provided in Table \ref{tab:i}, and how it compares to the 0-flavor RTA$^*$ curve.

$N$-component, i.e., $N$-species expansions on the shear viscosity have been previously addressed using various methods \cite{degroot, nasa, suth, Moroz:2014, MacKay:2024apd}. For asymptotically hot, i.e., ultra-relativistic anisotropic scatterings, the CE method was recently addressed \cite{MacKay:2024apd}. Specific useage of the traditional Anderson-Witting RTA viscosity for $N$-components was addressed in  Ref. \cite{Itakura:2007mx}, where a binary ($N=2$) hadronic gas (i.e., using FD statistics) was analyzed. However, the binary viscosity was treated as a linear combination of the ``pure gas'' viscosities of the components, and thus neglected the viscosity that would be produced by cross-species interference.

Naively, one can address $N$-component RTA$^*$ in the similar treatment as $N$-component RTA in the Anderson-Witting form (c.f. Eq. (59) in Ref. \cite{Chakraborty:2010fr}), however with the so-called molar fraction $x_i=n_i/\sum_{j=1}^N n_j$ \cite{degroot, Moroz:2014, MacKay:2024apd} that defines $\sum_{i=1}^N x_i=1$. For the specific case of massless MB-distributed partons with $\epsilon_a=3n_aT$,
\begin{equation} \label{nexprta}
\begin{split}
&\eta^{\mathrm{RTA^*}}_{~~~~N}=\frac{4T}{5}\sum_{a=1}^N x_a\tau^*_an_a,\\
&\mathrm{where}~~~\tau^*_a=\sum_{b=1}^N\frac{x_b}{n_b\langle\sigma^{ab}_{\mathrm{tr}}v_{\mathrm{rel}} \rangle}
\end{split}
\end{equation}
and $\langle\sigma^{ab}_{\mathrm{tr}}v_{\mathrm{rel}} \rangle$ is the thermal average in the form of Eq. (\ref{vrelint}) that includes the transport cross section for a purely elastic $a+b\rightarrow a+b$ interaction. The procedure would be to expand $\tau_a^*$ first, and then expand $\eta^{\mathrm{RTA^*}}_{~~~~N}$. 

 One can easily recover the pure gas ($N=1$) formula as Eq. (\ref{modrta}), and $N>1$ expansion is able to consider cross-species interactions. Just as the 0-flavor SU(3) comparison utilized the pure gas viscosity for the gluon-gluon interaction, the $N$-component RTA$^*$ formula should be compared with the AMY $N_f$-flavor basis function calculations for $m_D/T>1$ and under relevant ranges of $m_D/T= g\sqrt{(3+N_f/2)/3}$ for a collider-produced QGP.

\section{Conclusion}

The AMY formalism for expressing the shear viscosity is widely used in pQCD for its specific treatment of gluons and (anti-)quarks kinematically and statistically. As kinetic transport analysis on a collider-produced quark-gluon plasma utilizes 0-flavor SU(3) theory (i.e., a gluon gas approximation with added quark degrees of freedom), various analytical methods were used to express its shear viscosity within a confined temperature window. Such methods include a modified version of the relaxation time approximation and the Chapman-Enskog method. In the historical context, asymptotically hot QCD plasma was approached by RTA with an accounted $q^2$ momentum transfer. As briefly shown in Section \ref{comp}, the comparison between the 0-flavor SU(3) AMY formalism and RTA$^*$ expression for shear viscosity demonstrated that the numerical AMY calculations using basis functions mutually agree with the analytical RTA$^*$ expression within the relevant temperature window for a collider-produced QGP.

 Discussions of the results are offered in Section \ref{disc}, which addressed a second comparison that included the CE method under the exact gluon-gluon amplitude, and extension of AMY vs. RTA$^*$ analysis for larger flavor numbers (i.e., introduction of different particle types). In respective order, the CE curve has the similar but enhanced profile as the numerical AMY results, enhanced roughly by a factor of 2. Consideration of additional partons in a multi-component ($N_f>0$) QGP supposes the $N$-component expansion of the viscosity expression. For distinctive comparison between numerical $N_f$-flavor SU(3) AMY and analytical RTA$^*$, $N$-component expansion of the RTA$^*$ expression involves series expansion on the modified relaxation time to consider like- and cross-species interactions, as provided as Eq. (\ref{nexprta}) in Section \ref{ncomp}.

Continued research and analysis on collider-produced QGP is certain from a theoretical/mathematical perspective. This is especially the case for analyzing collider-produced QGP as a multi-component system of various partons. This requires an $N$-component expansion on RTA$^*$, and perhaps further comparison with larger quark-flavor AMY basis function calculations for mutual agreement. However, as the CE expression for shear viscosity better expresses collider-produced QGP, the respective $N$-component expansion would serve at the reliable mathematical model for multi-component collider-produced QGP viscosity.

\begin{appendices}




\end{appendices}



\begin{thebibliography}{}

\bibitem{J.Adams:STAR}
J.~Adams \textit{et al.} [STAR], 
Nucl. Phys. A \textbf{757}, 102 (2005)
doi:10.1016/j.nuclphysa.2005.03.085

\bibitem{PHENIX:2004vcz}
K.~Adcox \textit{et al.} [PHENIX],
Nucl. Phys. A \textbf{757}, 184-283 (2005)
doi:10.1016/j.nuclphysa.2005.03.086
[arXiv:nucl-ex/0410003 [nucl-ex]].

 \bibitem{hage}
M. Ga\'zdzicki, M.I. Gorenstein, M.I. ``Hagedorn’s Hadron Mass Spectrum and the Onset of Deconfinement.'' In: J. Rafelski (eds). \textit{Melting Hadrons, Boiling Quarks - From Hagedorn Temperature to Ultra-Relativistic Heavy-Ion Collisions at CERN} (Springer Publishers, Cham, 2016). Crossref: https://doi.org/10.1007/978-3-319-17545-4\_11
 
\bibitem{Romatschke:2007mq}
P.~Romatschke and U.~Romatschke,
Phys. Rev. Lett. \textbf{99}, 172301 (2007)
doi:10.1103/PhysRevLett.99.172301
[arXiv:0706.1522 [nucl-th]].
 
\bibitem{Song:2008si}
H.~Song and U.~W.~Heinz,
Phys. Rev. C \textbf{78}, 024902 (2008)
doi:10.1103/PhysRevC.78.024902
[arXiv:0805.1756 [nucl-th]].
 
\bibitem{ALICE:2010suc}
K.~Aamodt \textit{et al.} [ALICE],
Phys. Rev. Lett. \textbf{105}, 252302 (2010)
doi:10.1103/PhysRevLett.105.252302
[arXiv:1011.3914 [nucl-ex]].

\bibitem{Lin:2001zk}
Z.~w.~Lin and C.~M.~Ko,
Phys. Rev. C \textbf{65}, 034904 (2002)
doi:10.1103/PhysRevC.65.034904
[arXiv:nucl-th/0108039 [nucl-th]].

\bibitem{Xu:2007jv}
Z.~Xu, C.~Greiner and H.~Stocker,
Phys. Rev. Lett. \textbf{101}, 082302 (2008)
doi:10.1103/PhysRevLett.101.082302
[arXiv:0711.0961 [nucl-th]].

\bibitem{Xu:2008av}
Z.~Xu and C.~Greiner,
Phys. Rev. C \textbf{79}, 014904 (2009)
doi:10.1103/PhysRevC.79.014904
[arXiv:0811.2940 [hep-ph]].

\bibitem{Ferini:2008he}
G.~Ferini, M.~Colonna, M.~Di Toro and V.~Greco,
Phys. Lett. B \textbf{670}, 325-329 (2009)
doi:10.1016/j.physletb.2008.10.062
[arXiv:0805.4814 [nucl-th]].

\bibitem{Kovtun:2004de}
P.~Kovtun, D.~T.~Son and A.~O.~Starinets,
Phys. Rev. Lett. \textbf{94}, 111601 (2005)
doi:10.1103/PhysRevLett.94.111601
[arXiv:hep-th/0405231 [hep-th]].


\bibitem{Xu:2007ns}
Z.~Xu and C.~Greiner,
Phys. Rev. Lett. \textbf{100}, 172301 (2008)
doi:10.1103/PhysRevLett.100.172301
[arXiv:0710.5719 [nucl-th]].

\bibitem{Xu:2011fi}
J.~Xu and C.~M.~Ko,
Phys. Rev. C \textbf{83}, 034904 (2011)
doi:10.1103/PhysRevC.83.034904
[arXiv:1101.2231 [nucl-th]].

\bibitem{Plumari:2012ep}
S.~Plumari, A.~Puglisi, F.~Scardina and V.~Greco,
Phys. Rev. C \textbf{86}, 054902 (2012)
doi:10.1103/PhysRevC.86.054902
[arXiv:1208.0481 [nucl-th]].
 
\bibitem{MacKay:2022uxo}
N.~M.~MacKay and Z.~W.~Lin,
Eur. Phys. J. C \textbf{82}, no.10, 918 (2022)
doi:10.1140/epjc/s10052-022-10892-y
[arXiv:2208.06027 [nucl-th]].

\bibitem{Molnar:2001ux}
D.~Molnar and M.~Gyulassy,
Nucl. Phys. A \textbf{697}, 495-520 (2002)
[erratum: Nucl. Phys. A \textbf{703}, 893-894 (2002)]
doi:10.1016/S0375-9474(01)01224-6
[arXiv:nucl-th/0104073 [nucl-th]].
 
 \bibitem{ce}
 S. Chapman, T.G. Cowling. \textit{The Mathematical Theory of Non-Uniform Gases} (Cambridge University Press, Cambridge, 1995).

\bibitem{degroot}
S.R. de Groot, W.A. van Leeuwen, Ch. G. van Weert. \textit{Relativistic Kinetic Theory: Principles and Applications} (Amsterdam, 1980).

\bibitem{Anderson:1974nyl}
J.~L.~Anderson and H.~R.~Witting,
Physica \textbf{74}, no.3, 466-488 (1974)
doi:10.1016/0031-8914(74)90355-3

\bibitem{Wiranata:2012br}
A.~Wiranata and M.~Prakash,
Phys. Rev. C \textbf{85}, 054908 (2012)
doi:10.1103/PhysRevC.85.054908
[arXiv:1203.0281 [nucl-th]].

 \bibitem{Israel:1963}
W.~Israel,
 J. Math. Phys. \textbf{4}, 1163 (1963)
 doi:10.1063/1.1704047

\bibitem{Stewart:1971}
J.~M.~Stewart. \textit{Non-equilibrium Relativistic Kinetic Theory} (Springer-Verlag, Berlin, 1971).

\bibitem{Huovinen:2008te}
P.~Huovinen and D.~Molnar,
Phys. Rev. C \textbf{79}, 014906 (2009)
doi:10.1103/PhysRevC.79.014906
[arXiv:0808.0953 [nucl-th]].


\bibitem{Lin:2014tya}
Z.~W.~Lin,
Phys. Rev. C \textbf{90}, no.1, 014904 (2014)
doi:10.1103/PhysRevC.90.014904
[arXiv:1403.6321 [nucl-th]].
 
\bibitem{Zhang:1997ej}
B.~Zhang,
Comput. Phys. Commun. \textbf{109}, 193-206 (1998)
doi:10.1016/S0010-4655(98)00010-1
[arXiv:nucl-th/9709009 [nucl-th]].
 
\bibitem{Wang:2021owa}
H.~S.~Wang, G.~L.~Ma, Z.~W.~Lin and W.~j.~Fu,
Phys. Rev. C \textbf{105}, no.3, 034912 (2022)
doi:10.1103/PhysRevC.105.034912
[arXiv:2102.06937 [nucl-th]].

\bibitem{Arnold:2003zc}
P.~B.~Arnold, G.~D.~Moore and L.~G.~Yaffe,
JHEP \textbf{05}, 051 (2003)
doi:10.1088/1126-6708/2003/05/051
[arXiv:hep-ph/0302165 [hep-ph]].

\bibitem{mackay1}
N.~M.~MacKay. \textit{The shear viscosity of quark-gluon plasma under anisotropic scatterings} (Master's Thesis, East Carolina University, 2022).

\bibitem{Arnold:2000dr}
P.~B.~Arnold, G.~D.~Moore and L.~G.~Yaffe,
JHEP \textbf{11}, 001 (2000)
doi:10.1088/1126-6708/2000/11/001
[arXiv:hep-ph/0010177 [hep-ph]].
 
\bibitem{Thoma:1991em}
M.~H.~Thoma,
Phys. Lett. B \textbf{269}, 144-148 (1991)
doi:10.1016/0370-2693(91)91466-9

\bibitem{Heiselberg:1994vy}
H.~Heiselberg,
Phys. Rev. D \textbf{49}, 4739-4750 (1994)
doi:10.1103/PhysRevD.49.4739
[arXiv:hep-ph/9401309 [hep-ph]].

\bibitem{Baym:1990uj}
G.~Baym, H.~Monien, C.~J.~Pethick and D.~G.~Ravenhall,
Phys. Rev. Lett. \textbf{64}, 1867-1870 (1990)
doi:10.1103/PhysRevLett.64.1867

\bibitem{Baym:1991qf}
G.~Baym, H.~Monien, C.~J.~Pethick and D.~G.~Ravenhall,
Nucl. Phys. A \textbf{525}, 415C-418C (1991)
doi:10.1016/0375-9474(91)90355-A

\bibitem{Gross:1973id}
D.~J.~Gross and F.~Wilczek,
Phys. Rev. Lett. \textbf{30}, 1343-1346 (1973)
doi:10.1103/PhysRevLett.30.1343



\bibitem{Lin:2004en}
Z.~W.~Lin, C.~M.~Ko, B.~A.~Li, B.~Zhang and S.~Pal,
Phys. Rev. C \textbf{72}, 064901 (2005)
doi:10.1103/PhysRevC.72.064901
[arXiv:nucl-th/0411110 [nucl-th]].

\bibitem{schwartz}
M.D. Schwartz. \textit{Quantum Field Theory and the Standard Model} (Cambridge University Press, 2014).

\bibitem{Csernai:2006zz}
L.~P.~Csernai, J.~I.~Kapusta and L.~D.~McLerran,
Phys. Rev. Lett. \textbf{97}, 152303 (2006)
doi:10.1103/PhysRevLett.97.152303
[arXiv:nucl-th/0604032 [nucl-th]].

\bibitem{griffiths}
D. Griffiths. \textit{Introduction to Elementary Particle Physics} (John Wiley \& Sons, 1987).



\bibitem{Chakraborty:2010fr}
P.~Chakraborty and J.~I.~Kapusta,
Phys. Rev. C \textbf{83}, 014906 (2011)
doi:10.1103/PhysRevC.83.014906
[arXiv:1006.0257 [nucl-th]].

\bibitem{Bhatnagar:1954zz}
P.~L.~Bhatnagar, E.~P.~Gross and M.~Krook,
Phys. Rev. \textbf{94}, 511-525 (1954)
doi:10.1103/PhysRev.94.511

\bibitem{Koch:1986ud}
P.~Koch, B.~Muller and J.~Rafelski,
Phys. Rept. \textbf{142}, 167-262 (1986)
doi:10.1016/0370-1573(86)90096-7

\bibitem{Ghiglieri:2018dib}
J.~Ghiglieri, G.~D.~Moore and D.~Teaney,
JHEP \textbf{03}, 179 (2018)
doi:10.1007/JHEP03(2018)179
[arXiv:1802.09535 [hep-ph]].

\bibitem{nasa}
 R.~S.~Brokaw. \textit{Viscosity of gas mixtures} (NASA Lewis Research Center, Cleveland, 1968).
 
 \bibitem{suth}
W.~Sutherland,
 Phil. Mag. \textbf{40} (1895), pp. 421-431.

\bibitem{Moroz:2014}
O.~Moroz, 
Comput. Fluids \textbf{90}, 9 (2014)
doi:10.1016/j.compfluid.2013.11.014
[arXiv:1402.2240 [physics.flu-dyn]]

\bibitem{MacKay:2024apd}
N.~M.~MacKay,
[arXiv:2406.07764 [nucl-th]].

\bibitem{Itakura:2007mx}
K.~Itakura, O.~Morimatsu and H.~Otomo,
Phys. Rev. D \textbf{77}, 014014 (2008)
doi:10.1103/PhysRevD.77.014014
[arXiv:0711.1034 [hep-ph]].

\end{thebibliography}
\end{document}